# Magnetotransport evolution and nonlinear Hall effect in altermagnetic MnTe

*Wei Zhou[1#], Zhifeng Xue[1#], Yunxing Li[1], Nannan Tang[1], Ye Tao[1], Dingyong Zhong[1], Jiawei Luo[2*], Donghui Guo[1], Huichao Wang[1*]*

[1]Guangdong Provincial Key Laboratory of Magnetoelectric Physics and Devices, Center for Neutron Science and Technology, School of Physics, Sun Yet-sen University, Guangzhou, China

[2]State Key Laboratory of Quantum Functional Materials & ShanghaiTech Laboratory for Topological Physics, School of Physical Science and Technology, ShanghaiTech University, Shanghai, 201210, China.

**ABSTRACT**- Hexagonal MnTe is a prototypical semiconducting altermagnet whose properties are heavily influenced by intrinsic disorder, yet how the resulting diverse transport regimes shape its magnetotransport behavior remains to be clarified alongside the role of relativistic spin-orbit coupling (SOC). Here, we present a systematic study of the anisotropic magnetoresistance (AMR), planar Hall effect (PHE), and nonlinear transport in MnTe bulk single crystals. Below the Néel temperature ($T_N \sim 304$ K), the emergence of high-order harmonics in AMR and PHE within the high-temperature metallic regime reveals the interplay of magnetic order, crystalline symmetry, and SOC. At relatively lower temperatures, the disappearance of higher-order symmetries coincides with a transport crossover into the hopping conduction regime, suggesting that carrier localization diminishes the transport sensitivity to the Fermi-surface topology. In addition, we detect distinct second-order nonlinear Hall signals, providing evidence for a macroscopic inversion-asymmetric response in altermagnetic MnTe. Extending the investigations into the localized regime provides key insights into the subtle role of disorder and SOC in macroscopic charge transport. Our work thus underscores the necessity of exploring magnetotransport across diverse conducting regimes to comprehensively understand altermagnetic properties.

# 1. Introduction

Altermagnets have emerged as an unconventional class of magnetic materials that combine the distinct advantages of ferromagnets (FM) and antiferromagnets (AFM) [1–5]. Characterized by a high spin polarization and vanishing net magnetization, altermagnets possess a non-relativistic spin splitting in momentum space that is fundamentally protected by the crystal symmetry. Among altermagnetic candidates, hexagonal α-MnTe stands out due to its high Néel temperature, semiconducting characteristics, and well-defined altermagnetic features. It has attracted significant attention as a compelling platform to explore this unique magnetic order [5–7] and novel electronic properties [8–12]. While the primary spin splitting is fundamentally a non-relativistic effect, the presence of relativistic spin-orbit coupling (SOC) inevitably influences the electronic structure [7,13–15] and physical properties [16,17]. Crucially, as a semiconductor, MnTe is susceptible to intrinsic defects and disorder [18-20], yet how the disorder-driven diverse conducting regimes modulate the transport behavior has remained elusive. Unveiling how these factors intertwine is essential for the deep understanding and practical application of altermagnetic spintronics.

Fundamentally, the magnetic order and crystalline symmetry of MnTe act as the primary factors dictating its magnetotransport behavior. For instance, recent symmetry-based analyses have established the phenomenological profiles of anisotropic magnetoresistance (AMR) and planar Hall effect (PHE) in MnTe [21]. While such features have been investigated in thin films [21-24], the intrinsic properties of bulk single crystals without influence of substrate strain or interface scattering as well as their evolution across different transport regimes remain unclear. On the other hand, although altermagnetic MnTe is traditionally described by a centrosymmetric NiAs-type structure belonging to the $D_{6h}$ point group without considering the magnetic order, recent theoretical and experimental studies have revealed a subtle non-centrosymmetric distortion in the lattice [25-27]. This symmetry reduction is a prerequisite for nonlinear phenomena [28–33]. This debate demonstrates the necessity of utilizing macroscopic nonlinear transport, a technique sensitive to the spatial inversion symmetry breaking, to scrutinize whether an effective inversion-asymmetric response manifests in this altermagnetic semiconductor.

In this work, we systematically investigate the magnetotransport behavior of altermagnetic MnTe bulk single crystals across different transport regimes. In the relatively high-temperature metallic regime, we observe a distinct six-fold symmetry in the AMR and a four-fold harmonic component in the PHE beyond the conventional two-fold term, both persisting up to near $T_N$. These high-order transport features reflect a joint manifestation of the crystalline symmetry and magnetic order with relativistic SOC. Upon cooling, the higher-order symmetries attenuate, leaving a two-fold symmetry. This crossover shows potential correlation with the charge transport shift from coherent band conduction to the hopping regime, where the carrier localization suppresses the transport sensitivity to the detailed Fermi-surface topology. Interestingly, the two-fold component remains robust within the localized regime, which can be attributed to the coupling between the underlying magnetic background and spin-orbit interactions. In addition, we observe a distinct second-order nonlinear Hall signal in this altermagnetic semiconductor, suggesting inversion-symmetry breaking in the system. Overall, our work underscores the necessity of exploring magnetotransport across diverse conducting regimes to comprehensively understand the coupling between magnetic order, SOC and symmetry in altermagnets.

## 2. Methods

Single crystals of MnTe were synthesized via the chemical vapor transport (CVT) method using iodine as the transport agent. Stoichiometric amounts of Mn (99.9%) and Te (99.99%) powders were sealed in an evacuated quartz tube with $I_2$ under high vacuum. The tube was heated to 820 °C (source zone) and 610 °C (growth zone) over 8 hours and held for one week, yielding crystals typically 1-3 mm wide and 0.1 mm thick. The crystalline phase was verified by X-ray diffraction (XRD) using Cu $K_\alpha$ radiation ($\lambda$=1.5406 Å). Elemental stoichiometry was examined using energy dispersive spectroscopy (EDS). Electron transport and magnetic characterizations were performed in a Physical Property Measurement System (PPMS DynaCool, Quantum Design) and a Magnetic Property Measurement System (MPMS, Quantum Design).

## 3. Results

The oriented out-of-plane XRD scan (Fig. S1(a)) of the MnTe bulk crystals exhibits well-defined peaks corresponding to the (002) and (004) planes, indicating an orientation along the $c$-axis. All diffraction peaks of the crushed powders in the XRD pattern can be well indexed to the hexagonal phase of MnTe (PDF No. 18-0814). In this structure (inset of Fig. 1(a)), Mn atoms form a hexagonal close-packed framework with Te atoms occupying the octahedral interstitial sites. The crystals show an atomic ratio of approximately 1:1 (Fig. S1(b)) and $p$-type conductivity (Fig. S1(c)), consistent with the characteristic properties of altermagnetic MnTe [34–38].

The temperature-dependent magnetization $\mathbf{M}(T)$ of MnTe under a field-cooling (FC) condition of 4000 Oe is presented in Fig. 1(a), which reveals a transition to an AFM state at ~300 K. Figure 1(b) displays the isothermal magnetization curves $\mathbf{M}(H)$. The d$\mathbf{M}$/d$\mathbf{H}$ analysis (inset of Fig. 1(b)) identifies the spin-flop field ($H_{sf}$) within the 0.2–2 T range [39], characterized by a low differential susceptibility below $H_{sf}$ that increases significantly above it. Notably, the enhancement of magnetization at low temperatures suggests the emergence of weak ferromagnetism, confirmed by the hysteresis loop in the $\mathbf{M}(H)$ data at 2 K (Fig. 1(c)). The remanent magnetization of approximately $3.3\times10^{-5}$ $\mu_B$/f.u. aligns with values reported for bulk α-MnTe crystals ($2.5\text{-}5.0\times10^{-5}$ $\mu_B$/Mn) [34–38].

The temperature-dependent resistivity of the MnTe single crystal measured from 350 K down to 2 K is shown in Fig. 1(d). A distinct anomaly in the resistivity profile appears near the Néel temperature ($T_N \approx 304$ K), originating from enhanced spin fluctuations associated with the magnetic phase transition. Within the 100–300 K range, the resistivity exhibits metallic-like conduction, which can be well reproduced by the phenomenological formula $\rho$ (T) = $\rho_0 + aT + bT^5$ (yellow dashed line in Fig. 1(d)) where $\rho_0$ denotes the residual resistivity. The fitting analysis demonstrates a dominant $T^5$ law with a subtle correction from the linear $T$ term. Theoretically, according to the Bloch-Grüneisen formula, pure acoustic phonon scattering yields a $T^5$ dependence at $T \ll \Theta_D$ [40], which characteristically linearizes into a $T$ dependence at elevated temperatures. Similarly, pure electron-magnon scattering requires T $\ll T_N$ to sustain a $T^5$ behavior [41]. Given that the Debye temperature $\Theta_D$ of MnTe is ~240 K [42] and the Néel temperature is $T_N \approx 304$ K, the 100–300 K range spans $0.42\text{-}1.25\Theta_D$ and 0.33-0.98 $T_N$. Consequently, the observed $T^5$ behavior cannot be solely attributed to a single scattering source. Instead, a comprehensive explanation of the resistivity profile needs to consider the complex interplay and overlapping of these scattering mechanisms.

Upon further cooling below 100 K, the resistivity deviates from the high-temperature fit and exhibits an upturn signifying the onset of carrier localization. The 20–100 K range marks a transport crossover regime between high-temperature itinerant conduction and the low-temperature deep localized state. With decreasing temperature, the carriers lose sufficient thermal energy to sustain extended band conduction and are heavily influenced by vacancy-induced defect states. Charge transport becomes dominated by the nearest-neighbor hopping (NNH) between localized sites within the impurity band. This phonon-assisted process requires a small thermal activation energy $\Delta$. The NNH mechanism can be quantitatively validated via Arrhenius fitting ($\ln\rho \sim \Delta/k_BT$) [43, 44]. Our linear fit $\ln\rho \sim \Delta/k_BT$ in this temperature range yields an averaged $\Delta$~1.4 meV (Fig. 1(e)). Crucially, $\Delta$ is orders of magnitude smaller than the intrinsic bandgap of MnTe (~1.4 eV), ruling out band-to-band thermal activation and confirming an impurity-band-dominated NNH process. Depending on the specific sample, the upper temperature bound for this NNH regime varies from 60 K to 100 K. Nevertheless, the data below 100 K manifest a pronounced departure from the empirical metallic behavior, indicating that the hopping mechanism has taken dominance.

As the temperature further drops below 20 K, a much steeper resistivity upturn is observed, marking the entry into a strongly localized state. The charge carriers can only move via phonon-assisted long-range tunneling between localized states nearest to the Fermi level. This behavior is well-described by the three-dimensional Mott variable-range hopping (VRH) model [45], following the characteristic temperature dependence $\ln\rho \sim T^{-1/4}$ (Fig. 1(f)). The $T^{-1/4}$ law underscores the dominant role of disorder-induced localization in governing the low-temperature ground state of MnTe, establishing a solid transport background for understanding the subsequent AMR/PHE variations.

In-plane angular-dependent AMR and PHE measurements were conducted and the raw signals were processed to eliminate geometric asymmetry [46,47]. Figures 2(a)-(c) present the temperature-dependent angular sweeps of the AMR and PHE of MnTe measured at 9 T for current along the $[2\bar{1}\bar{1}0]$ ($x$ axis). It is found that the AMR and PHE at relatively high temperatures exhibit a complex angular dependence that deviates from the conventional two-fold symmetry (also see Fig. S2). When the current direction was aligned along the $[01\bar{1}0]$ direction ($y$ axis) instead of $[2\bar{1}\bar{1}0]$ ($x$ axis), the two-fold AMR and PHE are also observed with the correction of high-order symmetric contribution (Figs. 2(d)-2(f)). Our MnTe single crystals are utilized in their as-grown state and inherently host multiple domain configurations. However, the magnitudes of $\sigma_{xx}$ and $\sigma_{yy}$ show differences as reported [21], which may be influenced by geometric factors, spatial variations in contact resistances, and the possible symmetry discussed later.

To quantify the components, we employed an empirical model expressed as a sum of $A_n\cos(n\theta+\alpha_n)$ [21,24], where $A_n$ and $\alpha_n$ correspond to the amplitude and phase offset of the $n$-th component with respect to the rotating angle $\theta$ of the in-plane magnetic field relative to the $x$ axis. Theoretically, symmetry-based analyses establish that 2nd, 4th, and 6th harmonic components for the AMR, and 2nd, 3rd, and 4th components for the PHE, are symmetry-allowed [21]. Executing a full-parameter fitting across these harmonics would lead to overfitting. To ensure a well-constrained analysis, we further restricted our fitting to the primary 2nd and 6th harmonic components for the AMR, and the 2nd and 4th components for the PHE due to the absence of any AHE signatures related to A3 in our samples.

The experimental data can be well fitted by this model (Figs. 2(a)-(f)). Conventionally, the 2-fold term depends on the relative orientation of the Néel vector with respect to the current, corresponds to the non-crystalline component commonly found in all magnetically ordered materials, i.e., even in polycrystalline ones. The 4-fold and 6-fold stems from the anisotropy of the Fermi surface as demonstrated by the angle-resolved photoemission spectroscopy results [5,7]. In the transverse PHE, symmetry analysis reveals that the 6-fold component is forbidden and the higher-order crystalline anisotropy is projected into the transverse channel as a 4-fold modulation. Notably, the fitting amplitudes $A_i$ exhibits a non-monotonic temperature dependence, characterized by an initial increase followed by a subsequent decrease at low temperatures (Figs. 2(g)-(h)). Upon warming above 300 K, these anisotropic signals vanish in the paramagnetic state. In addition, the field-dependent measurements reveal a saturation of both AMR and PHE amplitudes above 2 T (Fig. S3), aligning with the completed field-induced spin-flop process. These results reveal the crucial role of the magnetic order in these angular-dependent magnetotransport.

Prior magnetotransport studies on altermagnetic MnTe were restricted to epitaxial thin films [21-24]. Transport coefficients in thin films are influenced by extrinsic factors such as interfacial strain and scattering, and the AMR amplitudes vary depending on sample quality. This variation combined with strong temperature dependence precludes a direct or meaningful quantitative comparison of the absolute values of $A_n$ to those of our bulk crystals. Nevertheless, we observe that the six-fold component becomes more pronounced in the high-temperature metallic regime, which aligns with earlier reports where distinct six-fold symmetries were resolved within the metallic transport window. Crucially, we would like to emphasize that our work investigates high-quality bulk crystals and tracks the temperature-dependent evolution of these harmonic symmetries across multiple distinct transport regimes.

We further explore the second-order nonlinear Hall effects to study the possible non-centrosymmetric distortion in MnTe. The current-voltage characteristics across different temperatures were systematically evaluated under zero magnetic field. In the linear transport regime, the longitudinal voltage $V^{\omega}$ maintains Ohmic relationship with the input AC current $I^{\omega}$ (Figs. 3a and c), with its slope evolving in accordance with the $\rho$-$T$ curve. Notably, a pronounced second-harmonic transverse response emerges for current applied along both the $x$- and $y$-axes, confirming a robust nonlinear transport effect (Figs. 3b and d). This behavior is frequency independent (Fig. S4) and shows distinct anisotropy for driving current aligned along different crystalline axes (Fig. S5), thereby ruling out Joule heating or parasitic capacitive coupling and confirming their electronic origin. The linear $I^{\omega}$-$V^{\omega}$ feature provides evidence that our measurements are performed within the weak-field linear transport regime, where any field-induced inherent hopping non-linearities are negligibly weak and cannot account for the 2ω signal.

The standard analysis for nonlinear Hall transport intrinsically relies on a well-defined quasiparticle relaxation time within the itinerant conduction regime [2-33]. In our system, the nonlinear transport predominantly manifests within the localization regime. In this regime, the conduction is dominated by phonon-assisted hopping rather than band-like scattering. Consequently, the semi-classical Boltzmann approach is no longer rigorously valid. Although a quantitative analysis remains unachievable within this hopping framework, we plot the temperature dependence of the slopes extracted from both the first-order and second-order Hall responses to facilitate a phenomenological understanding (Figs. 4(a)-(b), 4(d)-(e)). Furthermore, to eliminate potential artifacts arising from sample geometric dimensions and thickness uncertainties, we have introduced a size-independent physical quantity directly related to the second-order conductivity tensor, the ratio of the second-order current-voltage ($I^{\omega 2}$-$V^{2\omega}$) slope to the cube of the first-order ($I^{\omega}$-$V^{\omega}$) slope. The evolution of this parameter as a function of temperature is presented in Figs. 4(c) and 4(f).

# 4. Discussion

The transport anisotropy in traditional metallic ferromagnets and antiferromagnets is generally understood by the SOC mediated anisotropic s-d scattering [48,49]. The AMR and PHE microscopically originate from the anisotropic electronic relaxation time modulated by the lattice. If SOC was turned off, the spin completely decouple from the lattice, resulting in an isotropic charge scattering rate and a vanishing AMR/PHE. Although recent theoretical work has proposed “non-relativistic AMR” mechanisms in certain systems [50], the requirement of breaking the real-space rotational symmetry is not satisfied in MnTe and such a response yields only a 2-fold symmetry. A recent study reveals a physical mechanism where a giant AMR response is initially driven by the non-relativistic spin splitting under a zero net magnetic moment background, yet the final macroscopic electrical readout still necessitates a finite SOC via spin-to-charge conversion [51]. The indispensable role of SOC is also confirmed by recent density functional theory (DFT) calculations for MnTe [52]. The large AMR/PHE primarily originates from the spin-orbit-coupled hole pockets along the Γ-K lines, whose Fermi surface geometry is highly sensitive to the Néel-vector orientation. On the other hand, at the A-top case, where spin properties are dominated by the nonrelativistic altermagnetic feature, the AMR is flat with much smaller oscillation. Therefore, the observation of AMR/PHE of MnTe in the metallic regime fundamentally requires the participation of SOC.

Regarding the suppression of the higher-order (4-fold and 6-fold) components at lower temperatures, we evaluated the potential correlation with the spin-reorientation around 81 K [11], which corresponds to an in-plane $\pi/2$ rotation of the easy magnetization axis. However, the periodicity of M oscillations at fixed magnetic fields across the transition temperature remains unchanged. The symmetry-preserving transition may cause a phase shift or modulation of transport tensor components, while it fundamentally lacks the mechanism to eliminate the higher-order harmonic terms of the AMR and PHE. In contrast, the temperature dependence coincides with the regime deviating from metallic behavior, suggesting a correlation with the transport mechanism transition. At high temperatures $T > 100$ K, transport is dominated by coherent band-like scattering. In this regime, the charge transport is highly sensitive to the fine geometric details of the Fermi surface. Thus, the crystalline anisotropy is clearly resolved, giving rise to 4-fold and 6-fold harmonics. Upon entering the localized transport regime, the clear Fermi-surface geometry dissolves, and conduction is instead mediated by hopping trajectories within a disordered percolation network. Within this random network, the long-range crystalline spatial coherence is lost, and the statistical averaging over random hopping events eliminates any high-order modulations. For the robust 2-fold anisotropy in the hopping regime, the coupling of localized transport to magnetic moments has been widely suggested [53-56]. Concurrently, a spin-orbit coupling background is required to dictate the transverse charge accumulation [53,57,58]. Therefore, we attribute the AMR and PHE in MnTe in the hopping regime to the effects of magnetic backgrounds and spin-orbit interactions, while a precise microscopic picture warrants further theoretical and experimental investigations.

The emergence of second-order Hall transport in MnTe provides evidence for an effective macroscopic inversion-asymmetric response, while transport measurements alone are insufficient to establish its microscopic origin. Though the idealized NiAs-type lattice suggests a centrosymmetric structure, crucial evidence from recent literature solidifies this foundation for the inversion-symmetry breaking in MnTe [25-27]. Specifically, the optical studies identify a noncentrosymmetric crystal symmetry [25], and direct atomic-scale imaging reveals inversion-symmetry-breaking distortions [26]. In addition, a recent theory provides a qualitative framework demonstrating that such contribution from the lattice polarization coexists with the nonlinear anomalous Hall effect in MnTe [27]. Together, these results provide indispensable support for our observed 2ω nonlinear Hall responses in altermagnetic MnTe.

To the best of our knowledge, a dedicated theoretical framework for the nonlinear Hall effect in the localized hopping regime remains a vacant and unexplored frontier. To provide a reasonable physical picture, we tentatively extend the established microscopic models of localized transport to our nonlinear observations from a qualitative perspective. Following the hopping transport model [53,57], the transverse transport can originate from a quantum phase acquired during hopping around closed-loop paths, enabled by spin-orbit interaction and magnetic background. Importantly, recent studies on hopping transport suggest that intrinsic spin splitting protected by crystal symmetry dictates pronounced transport anisotropy in altermagnets [55]. By projecting this into the hopping framework, we suggest the possibility that inherent features render localized tunneling rates spin-dependent and anisotropic, with SOC acting as a necessary mediator to translate these asymmetries into the nonlinear Hall effect. We expect these experimental results will serve as a timely catalyst, inspiring the community to develop models for nonlinear transport in the localized hopping regime. Meanwhile, further investigations into the role of SOC and its intricate interplay with topology in altermagnets are highly anticipated to enrich this emerging field [58,59].

In summary, we present the magnetotransport properties evolution of bulk MnTe single crystals and reveal how disorder-driven transport mechanisms influence the macroscopic manifestation. Within the itinerant metallic regime, the emergence of higher-order harmonic terms in the AMR and PHE reflects an intricate interplay between the altermagnetic order, crystal symmetry, and relativistic SOC. Upon cooling into the hopping regime, carrier localization suppresses the transport sensitivity to the Fermi-surface anisotropy, leading to the attenuation of higher-order anisotropic features while the fundamental two-fold term persists. In addition, we observe a distinct zero-field second-order nonlinear Hall effect, providing evidence for a macroscopic inversion-asymmetric response in this altermagnetic semiconductor. By extending the investigation of both linear anisotropy and nonlinear transport into the carrier-localized regime, our findings demonstrate the necessity of multi-regime transport exploration to comprehensively understand unconventional altermagnets.

## AUTHOR INFORMATION

### Corresponding Author

*wanghch26@mail.sysu.edu.cn

*luojw2@shanghaitech.edu.cn

### Author Contributions

†W. Z. and Z. F. X. contributed equally.

The authors declare no competing financial interest.

## ACKNOWLEDGMENT

We acknowledge the support from the National Natural Science Foundation of China (Nos. 12374052, 92565303), Guangdong Provincial Quantum Science Strategic Initiative (No. GDZX2401009), the Interdisciplinary program of Wuhan National High Magnetic Field Center (No. WHMFC2025017), Huazhong University of Science and Technology, Guangzhou Basic and Applied Basic Research Foundation (No.

2025A04J5405), Research Center for Magnetoelectric Physics of Guangdong Province (No. 2024B0303390001), Guangdong Provincial Key Laboratory of Magnetoelectric Physics and Devices (No. 2022B1212010008).

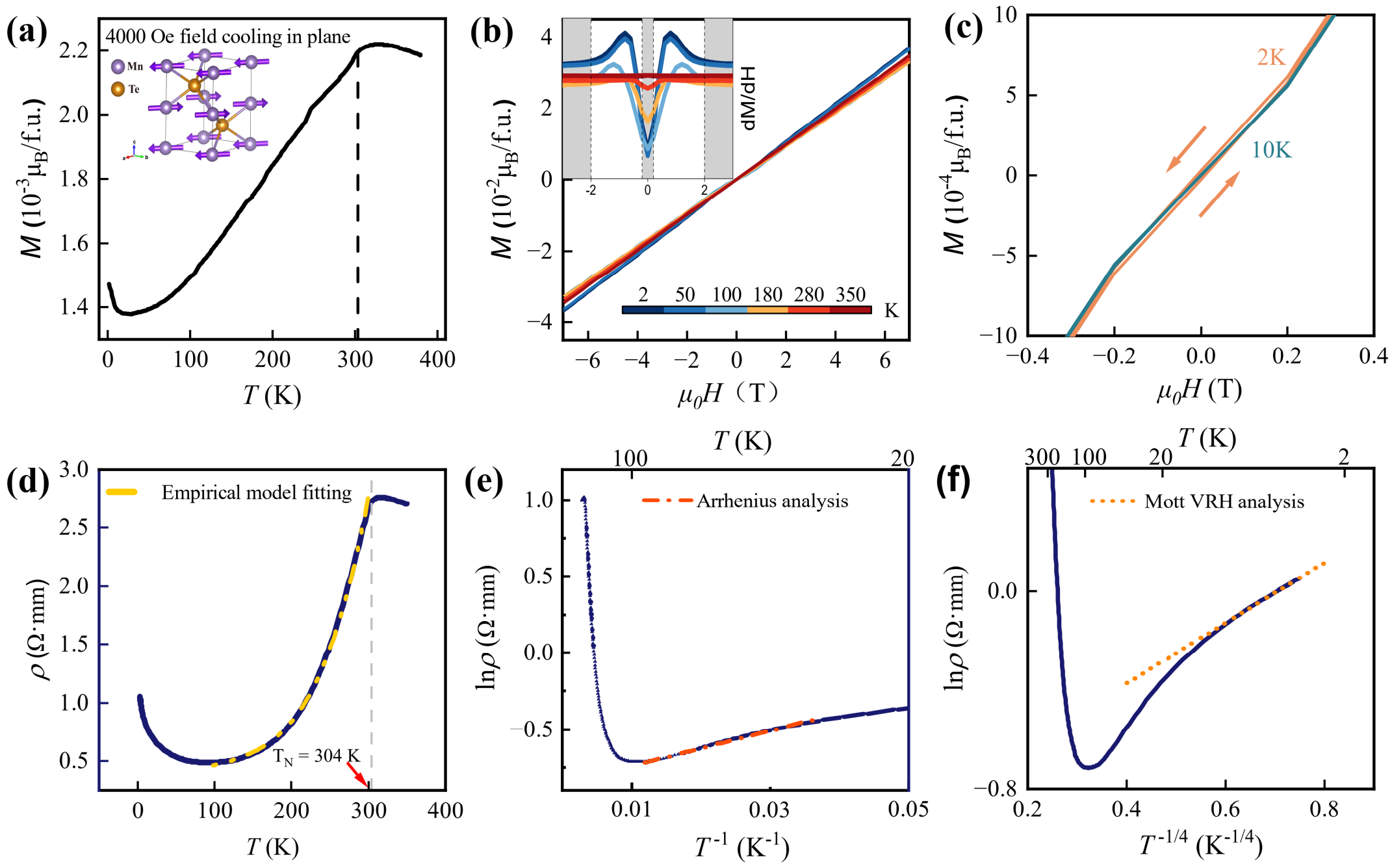

**Figure 1.** (a) Temperature dependence of magnetization measured under field-cooling at 4000 Oe. The inset shows a schematic of crystal structure. (b) Isothermal magnetization M(H) curves at various temperatures from 2 to 350 K. The inset shows the differential susceptibility dM/dH, identifying the spin-flop process in 0.2-2 T. (c) Low-temperature M(H) hysteresis loop. (d) Temperature dependence of the resistivity (sample 1). The yellow dashed line is a fitting using $\rho(T) = \rho_0 + aT + bT^5$, indicating high-temperature metallic conduction. (e) Arrhenius analysis ($\ln\rho \sim T^{-1}$) in the intermedium temperature range. The localization effect becomes dominant in this regime. (f) Plot of $\ln\rho$ versus $T^{-1/4}$ at low temperatures. The linear fit indicates that the charge transport in the regime is governed by three-dimensional Mott VRH mechanism.

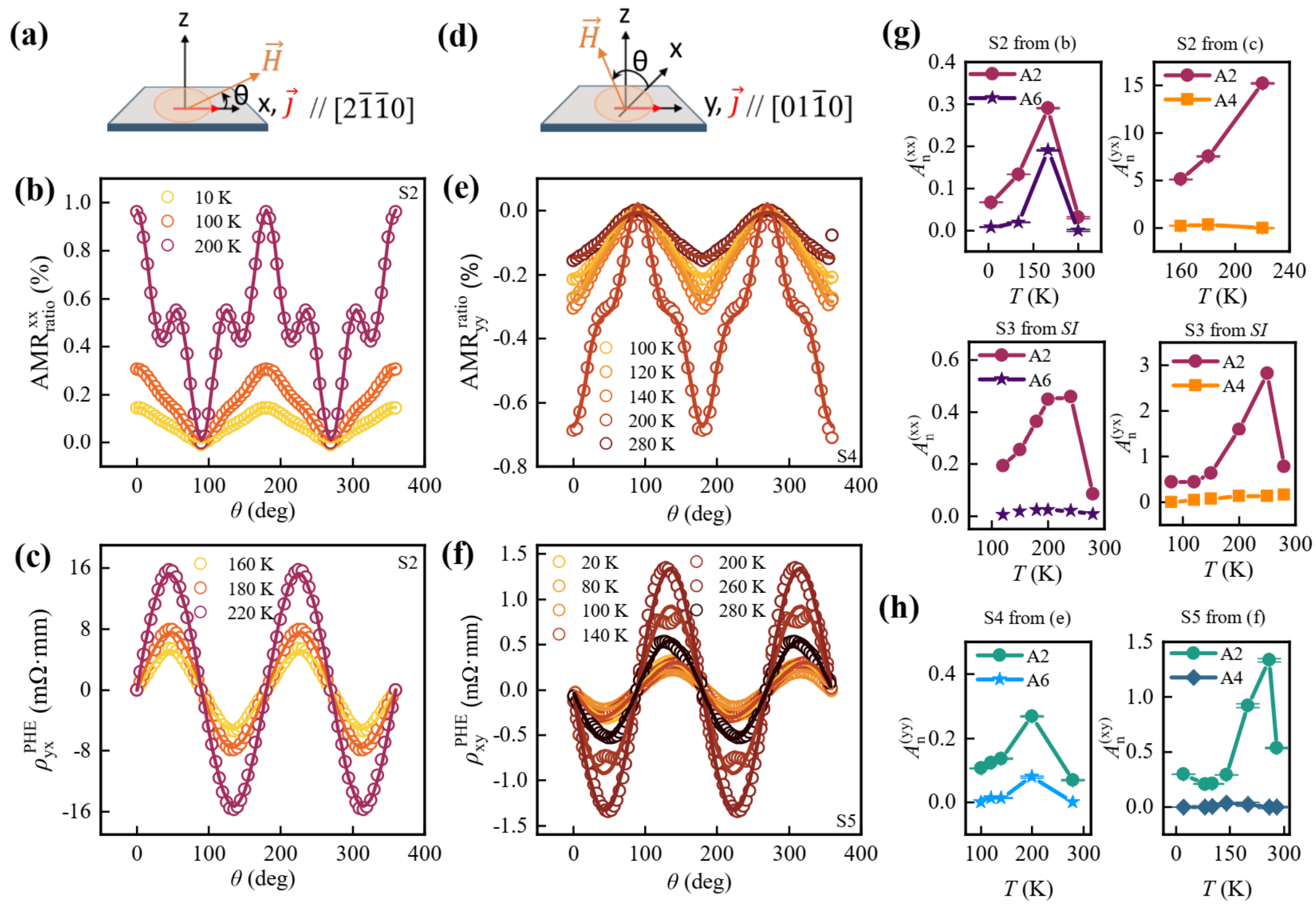

**Figure 2.** (a) Experimental schematic, (b) AMR and (c) PHE measured at various temperatures under a fixed 9 T with the current *j* applied along the [2$\bar{1}\bar{1}$0] direction (*x*-axis). (d) Experimental schematic, (e) AMR and (f) PHE measured at 9 T with the current *j* applied along the [01$\bar{1}$0] direction (*y*-axis). Here the AMR ratio is defined as $(\rho(\theta)-\rho(\pi/2))/\rho(\pi/2) \times 100\%$. Solid lines in (b,c,e,f) represent fits based on the phenomenological model expressed as a sum of $A_n\cos(n\theta+\alpha_n)$. An additional constant is also introduced for AMR. (g,h) Temperature dependence of the amplitudes $A_n$ obtained with corresponding data labeled. The uncertainties of the data points are represented by error bars.

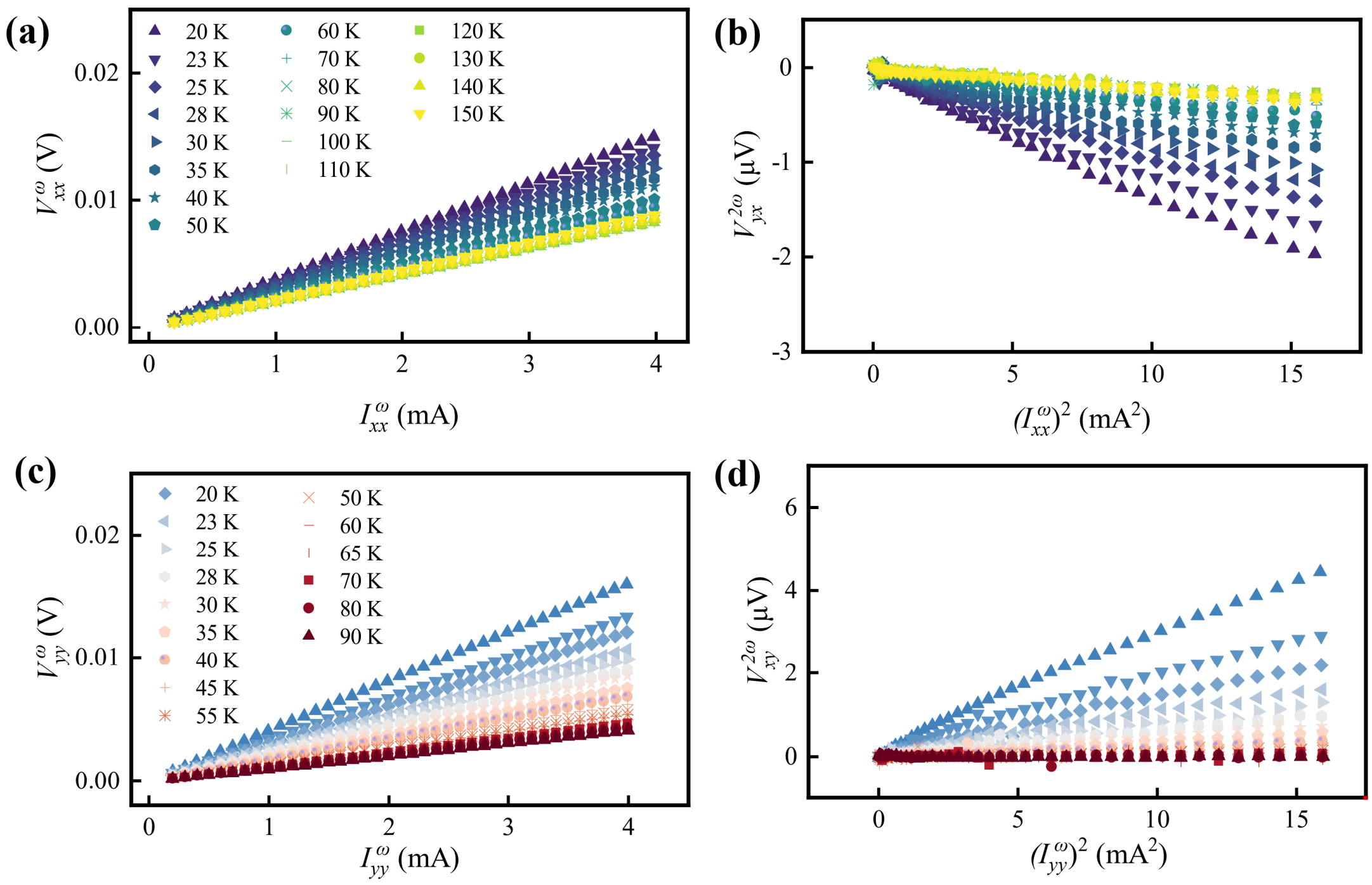

**Figure 3.** (a, c) Longitudinal voltages ($V_{xx}^{\omega}$ and $V_{yy}^{\omega}$) versus the AC current ($I^{\omega}$) measured under zero magnetic field along the *x*- and *y*-axes (sample 6), demonstrating linear Ohmic behavior. (b, d) Transverse second-harmonic voltages ($V_{yx}^{2\omega}$ and $V_{xy}^{2\omega}$) versus the square of the applied current. The linear scaling confirms an intrinsic quadratic transport response.

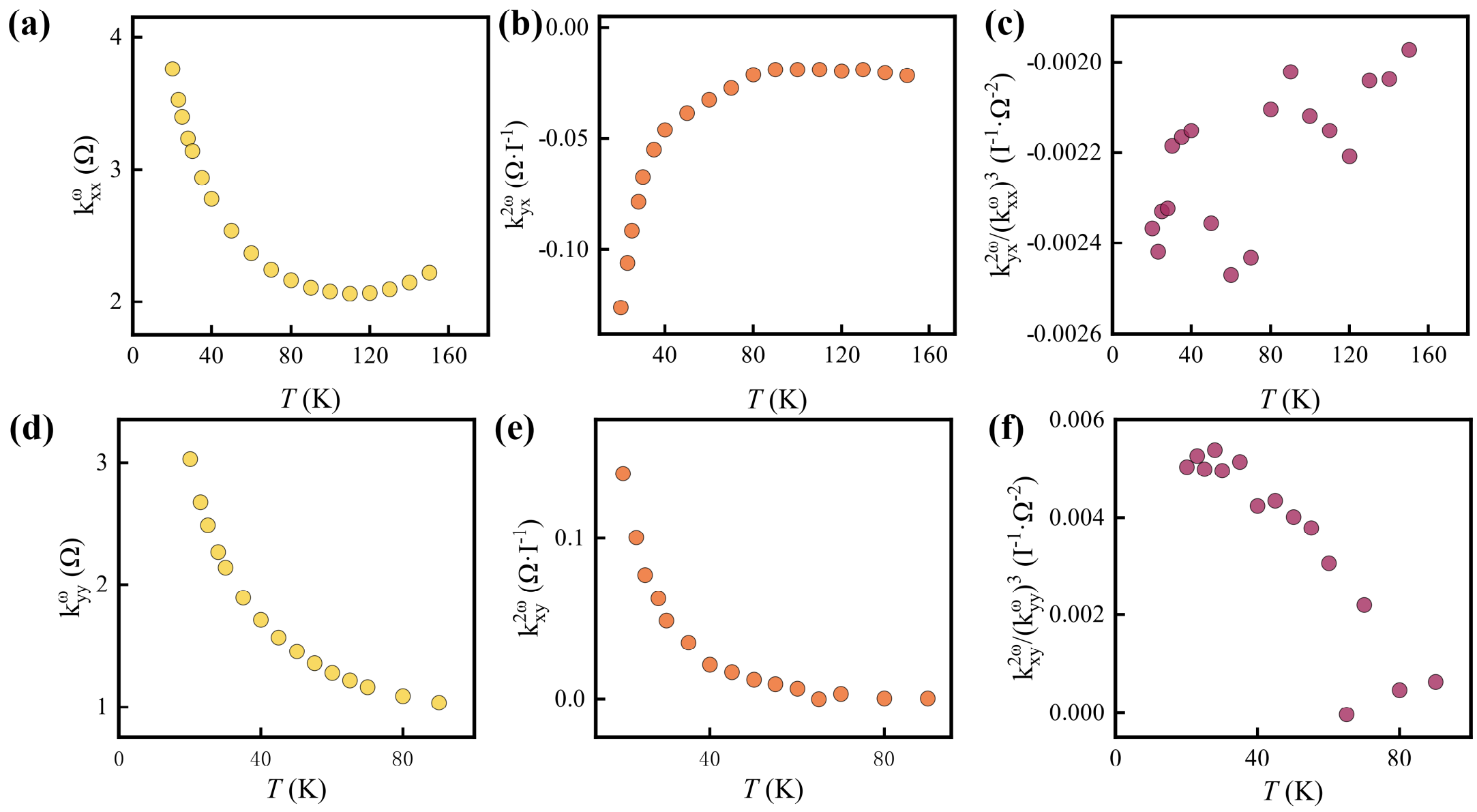

**Figure 4.** Temperature dependence of (a, d) first-harmonic coefficients (defined as $V^{\omega}/I^{\omega}$), (b, e) second-harmonic nonlinear coefficients (defined as $V^{2\omega}/I^{\omega 2}$), and (c, f) their ratios of the second-harmonic coefficient divided by the cube of the first-harmonic coefficient.